\documentclass[aps,pra,reprint,showpacs,floatfix,longbibliography]{revtex4-1}
\usepackage{graphicx, amsmath, amssymb}
\usepackage[caption=false]{subfig}
\allowdisplaybreaks[1] 
\begin{document}


\newcommand{\braket}[2]{{\left\langle #1 \middle| #2 \right\rangle}}
\newcommand{\bra}[1]{{\left\langle #1 \right|}}
\newcommand{\ket}[1]{{\left| #1 \right\rangle}}
\newcommand{\ketbra}[2]{{\left| #1 \middle\rangle \middle \langle #2 \right|}}


\title{Faster Quantum Walk Search on a Weighted Graph}

\author{Thomas G.~Wong}
	\email{twong@lu.lv}
	\affiliation{Faculty of Computing, University of Latvia, Rai\c{n}a bulv.~19, R\=\i ga, LV-1586, Latvia}

\begin{abstract}
	A randomly walking quantum particle evolving by Schr\"odinger's equation searches for a unique marked vertex on the ``simplex of complete graphs'' in time $\Theta(N^{3/4})$. In this paper, we give a weighted version of this graph that preserves vertex-transitivity, and we show that the time to search on it can be reduced to nearly $\Theta(\sqrt{N})$. To prove this, we introduce two novel extensions to degenerate perturbation theory:\ an adjustment that distinguishes the weights of the edges, and a method to determine how precisely the jumping rate of the quantum walk must be chosen.
\end{abstract}


\maketitle


\section{Introduction}

Grover's algorithm \cite{Grover1996} famously searches a ``database'' of size $N$ for a ``marked'' item by querying an oracle $O(\sqrt{N})$ times. This unstructured search assumes that there is no structure to the database, so one can move from querying any item to querying any other, and information about the location of the marked vertex can only come from querying the oracle. As such, it is natural to formulate Grover's algorithm as a quantum walk \cite{Kempe2003} on the complete graph of $N$ vertices \cite{FG1998a,CG2004,Wong2015b}, since every vertex is connected to every other, and the goal is to find a marked vertex by querying an oracle.

Physically, however, structure may exist that prevents one from arbitrarily traversing the database. Such spatial search problems can also be modeled as searches on graphs \cite{AA2005,CG2004,AKR2005}. Although spatial search algorithms using quantum walks have been around for roughly a decade, there is still no complete theory as to what structures support fast quantum search \cite{MeyerWong2015}.

\begin{figure}
\begin{center}
	\includegraphics{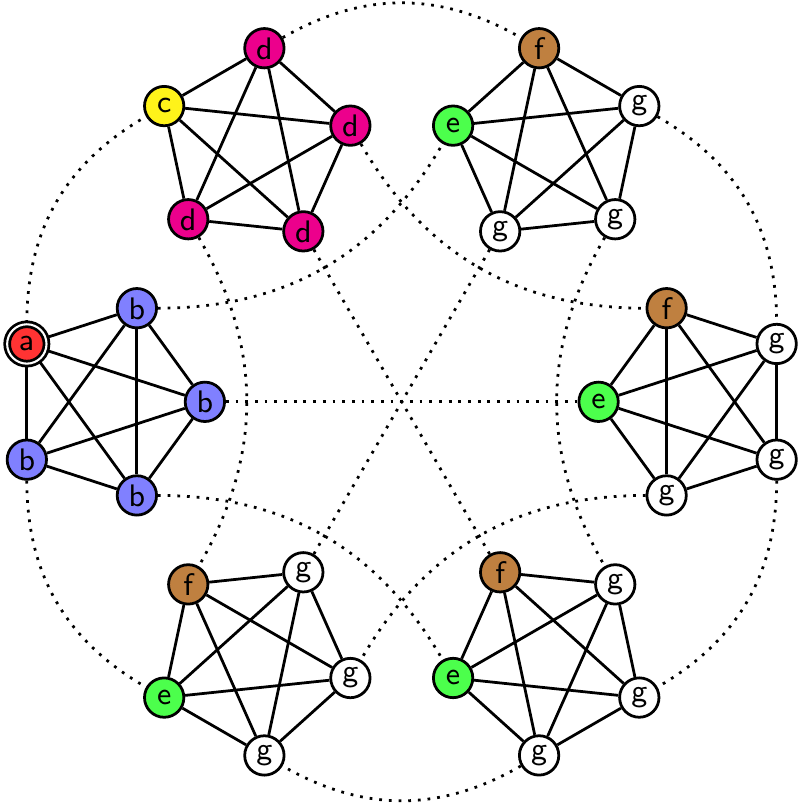}
	\caption{\label{fig:simplex_weighted} A 5-simplex with each vertex replaced by the complete graph of 5 vertices. Solid edges have weight $1$, and dotted edges have weight $w$. A vertex is marked, indicated by a double circle. Identically evolving vertices are identically colored and labeled.}
\end{center}
\end{figure}

One graph that has recently provided several new insights into the role of structure in spatial search is the ``simplex of complete graphs,'' an example of which is shown in Fig.~\ref{fig:simplex_weighted}. This graph is an arrangement of $M+1$ complete graphs of $M$ vertices such that each vertex in a complete graph is connected to a different cluster. Formally, it is a first-order truncated $M$-simplex lattice \cite{Dhar1977}, and it contains $N = M(M+1)$ vertices and $M^2(M+1)/2$ edges.

This graph has enough structure to yield interesting results, but enough symmetry to lend itself to analysis. It was first introduced for quantum search in \cite{MeyerWong2015} as a structure with high connectivity, but on which search is unexpectedly slow. It has also been studied with various spatial distributions of multiple marked vertices \cite{Wong2015a}. Finally, search for a completely marked cluster causes the standard continuous-time quantum walk search algorithm to beat the typical discrete-time one \cite{WongAmbainis2015}. In each of these, the graph was unweighted (\textit{i.e.}, each edge had weight $1$).

In this paper, we also focus on search on the simplex of complete graphs, except now we consider a weighted version where the edges within clusters still have weight $1$, but edges between clusters have weight $w \in \mathbb{R}_{+}$. This is denoted by the solid and dotted edges in Fig.~\ref{fig:simplex_weighted}, respectively. Although quantum walks on weighted graphs have been investigated for universal mixing \cite{CFHRTW2007} and utilized for quantum state transfer \cite{CDEL2004} and quantum transport \cite{ZFKWLB2013}, they have not been explored for search (apart from the context of time-reversal symmetry breaking \cite{Wong2015d}, which yielded no speedup).

With this choice of weights, the graph remains vertex-transitive, so regardless of which vertex is marked, the system will evolve the same. As we explain later, weights defined this way on the simplex of complete graphs preserves some key properties of quantum search, and importantly, we show that search can be faster on the the weighted graph. In particular, search for a unique marked vertex on the unweighted ($w = 1$) graph takes time $\Theta(N^{3/4})$ \cite{MeyerWong2015}. We show that as $w$ increases, the runtime decreases to $\Theta(N^{3/4}/w)$. We can choose $w$ to nearly scale as $N^{1/4}$, reducing the runtime to nearly $\Theta(\sqrt{N})$.

Next, we define the quantum walk search algorithm on the graph, and we show that the system evolves in a constant $7$-dimensional subspace. Then we explain the general evolution of the algorithm, which occurs in two stages. Afterwards we analytically prove the runtime of the algorithm, which involves novelly adjusting degenerate perturbation theory \cite{JMW2014} to capture the weights of the graph. Subsequently, we give a new method of using degenerate perturbation theory to determine how precisely the jumping rate of the quantum walk must be chosen. Since degenerate perturbation theory is a useful tool in a variety of quantum search problems \cite{JMW2014,MeyerWong2015,Wong2015a,WongAmbainis2015,Novo2014}, these two extensions to the method are important apart from the graph at hand. Finally, we end with some remarks about the energy usage of the search algorithm and the connectivity of the weighted graph.


\section{Quantum Walk Search}

The vertices of the graph label computational basis states of an $N$-dimensional Hilbert space. A randomly walking quantum particle searches for a ``marked'' vertex of a regular graph by evolving by Schr\"odinger's equation with Hamiltonian
\[ H = -\gamma A - \ketbra{a}{a}, \]
where $\gamma$ is the jumping rate (\textit{i.e.}, amplitude per time), $A$ is the adjacency matrix of the graph ($A_{ij}$ equals the weight of the edge between vertices $i$ and $j$, and is zero if they are not connected), and $\ket{a}$ is a vertex that we are searching for \cite{CG2004}. Together, $-\gamma A$ effects a quantum walk, and $-\ketbra{a}{a}$ is a Hamiltonian oracle \cite{Mochon2007}.

The system $\ket{\psi(t)}$ begins in an equal superposition $\ket{s}$ over all the vertices:
\[ \ket{\psi(0)} = \ket{s} = \frac{1}{\sqrt{N}} \sum_{i=1}^N \ket{i}. \]
Not only is this a convenient initial state, but it expresses our initial lack of knowledge of where the marked vertex might be by guessing each vertex with equal probability. It is also an eigenstate of the adjacency matrix $A$ (with eigenvalue $M+w-1$) that effects the quantum walk, so if we evolve by $-\gamma A$ alone, we have no new information, and the state stays the same. It is only when we include the oracle term $-\ketbra{a}{a}$ that information changes, and the state changes with it.

Figure ~\ref{fig:simplex_weighted} shows that there are only seven kinds of vertices. For example, all the blue $b$ vertices will evolve the same way. Thus we can group identically-evolving vertices together into a 7D subspace:
\begin{align*}
	\ket{a} &= \ket{\text{red}} \\
	\ket{b} &= \frac{1}{\sqrt{M-1}} \sum_{i \in \text{blue}} \ket{i} \\
	\ket{c} &= \ket{\text{yellow}} \\
	\ket{d} &= \frac{1}{\sqrt{M-1}} \sum_{i \in \text{magenta}} \ket{i} \\
	\ket{e} &= \frac{1}{\sqrt{M-1}} \sum_{i \in \text{green}} \ket{i} \\
	\ket{f} &= \frac{1}{\sqrt{M-1}} \sum_{i \in \text{brown}} \ket{i} \\
	\ket{g} &= \frac{1}{\sqrt{(M-1)(M-2)}} \sum_{i \in \text{white}} \ket{i}.
\end{align*}
In this subspace, the initial equal superposition state is
\begin{align*}
	\ket{s} = \frac{1}{\sqrt{N}} \Big( &\ket{a} + \sqrt{M-1} \ket{b} + \ket{c} + \sqrt{M-1} \ket{d} \\ &+ \sqrt{M-1} \ket{e} + \sqrt{M-1} \ket{f} \\ &+ \sqrt{(M-1)(M-2)} \ket{g} \Big),
\end{align*}
and the search Hamiltonian is
\[ \setlength{\arraycolsep}{1pt} H = -\gamma \! \left( \! \begin{matrix}
	\frac{1}{\gamma} & \sqrt{M_1} & w & 0 & 0 & 0 & 0 \\
	\sqrt{M_1} & M_2 & 0 & 0 & w & 0 & 0 \\
	w & 0 & 0 & \sqrt{M_1} & 0 & 0 & 0 \\
	0 & 0 & \sqrt{M_1} & M_2 & 0 & w & 0 \\
	0 & w & 0 & 0 & 0 & 1 & \sqrt{M_2} \\
	0 & 0 & 0 & w & 1 & 0 & \sqrt{M_2} \\
	0 & 0 & 0 & 0 & \sqrt{M_2} & \sqrt{M_2} & M_3+w \\
\end{matrix} \! \right) \! , \setlength{\arraycolsep}{3pt} \]
where $M_k = M - k$. Thus we have reduced an $N$-dimensional problem to a 7-dimensional one.


\section{Two-Stage Algorithm}

\begin{figure*}
\begin{center}
	\subfloat[]{
		\includegraphics{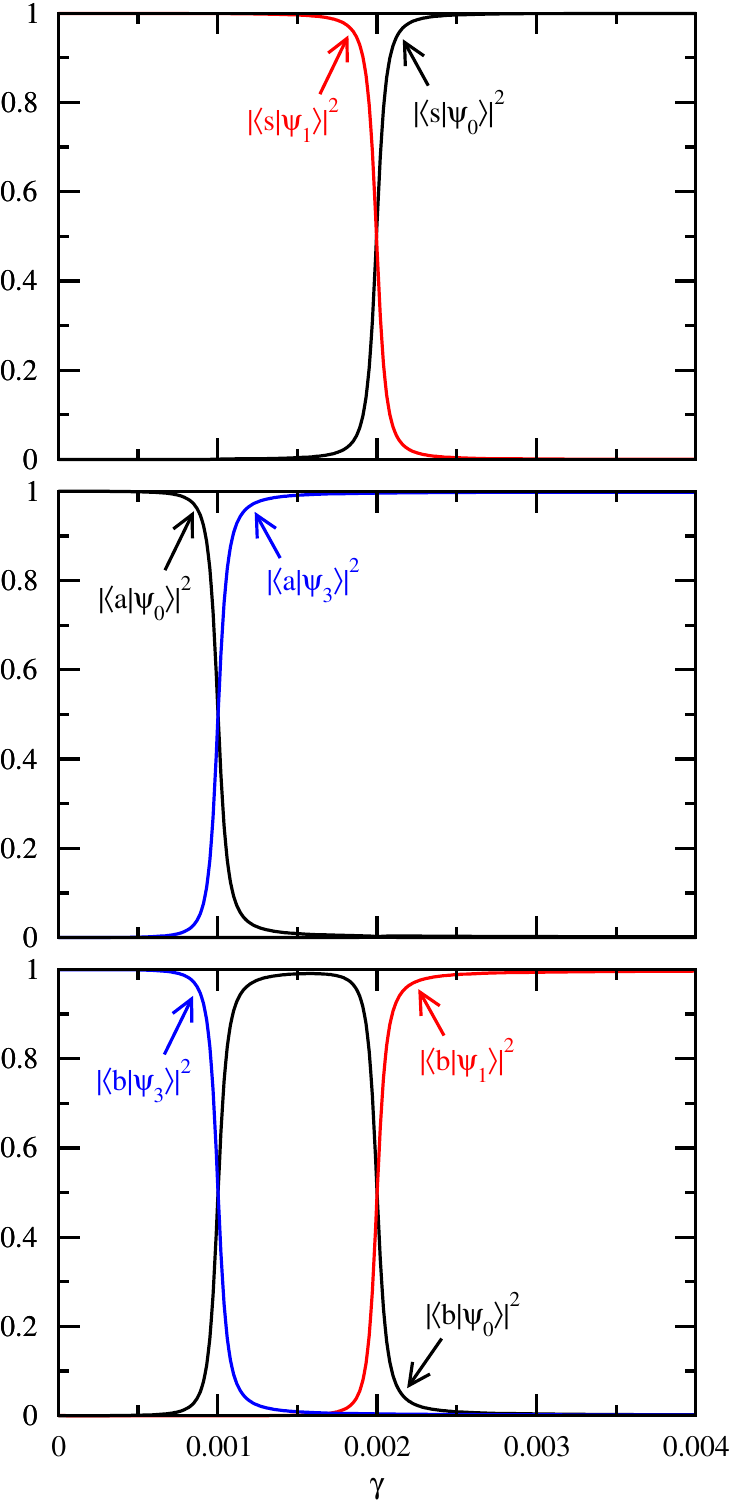}
		\label{fig:overlap_w1} 
	} \quad \quad \quad \quad
	\subfloat[]{
		\includegraphics{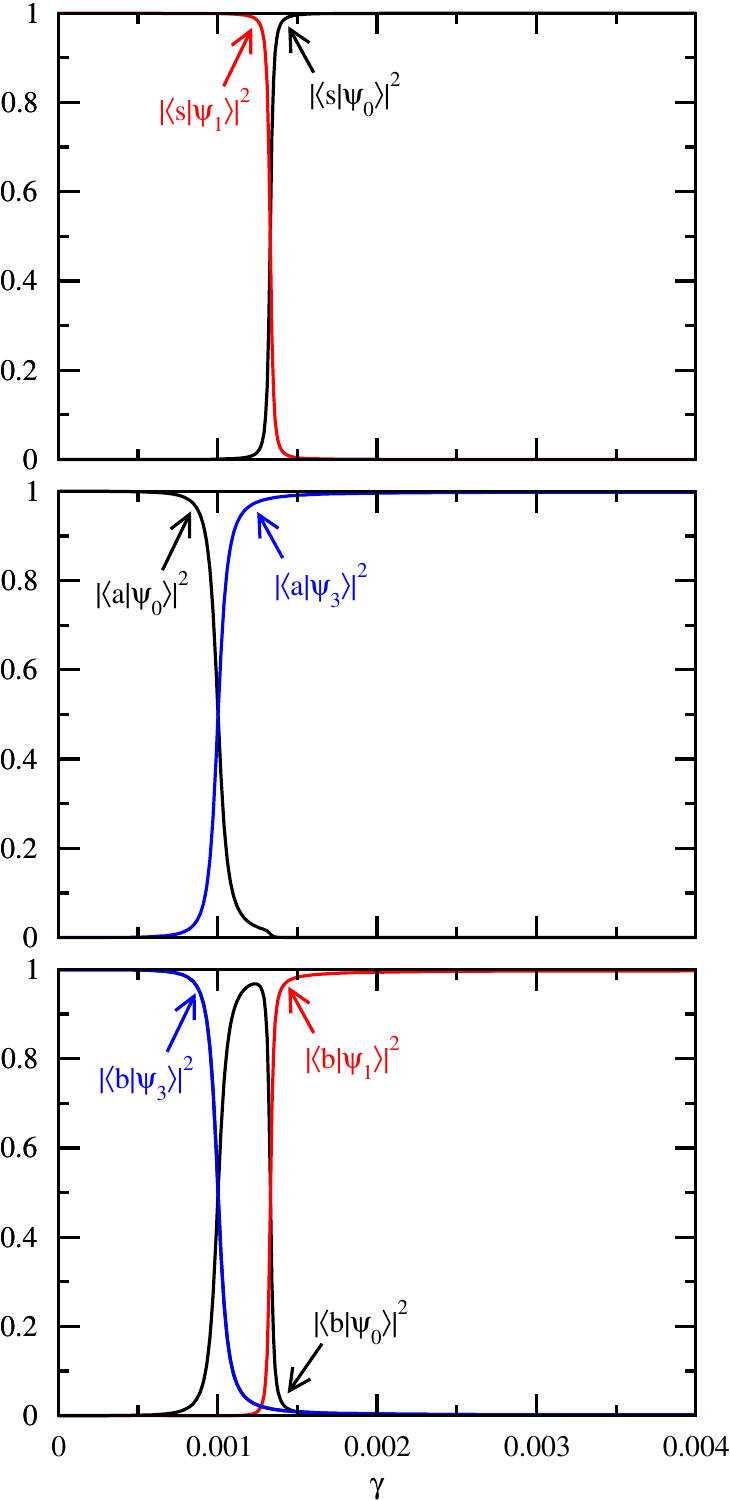}
		\label{fig:overlap_w3} 
	}
	\caption{\label{fig:overlap} Probability overlaps of $\ket{s}$, $\ket{a}$, and $\ket{b}$ with eigenstates of $H$ for search on the weighted simplex of complete graphs with $M = 1000$ and (a) $w = 1$ (unweighted), and (b) $w = 3$.}
\end{center}
\end{figure*}

To understand the behavior of the algorithm, let us start with the unweighted ($w = 1$) graph, which was first solved in \cite{MeyerWong2015}, and whose solution we now summarize. In Fig.~\ref{fig:overlap_w1}, we plot the probability overlaps of $\ket{s}$, $\ket{a}$, and $\ket{b}$ with the eigenvectors of $H$. When $\gamma$ is away from the critical value $\gamma_{c1} = 2/M = 0.002$, the initial state $\ket{s}$ asymptotically equals the ground or first excited state, and the system fails to evolve. So for the system to evolve at all, we must pick $\gamma$ to equal $\gamma_{c1}$, where the initial state takes the form $\ket{s} \propto \ket{\psi_0} + \ket{\psi_1}$. Thus it is half in the ground state and half in the first excited state. At $\gamma_{c1}$, note that $\ket{b}$ is also half in each of those energy eigenstates, taking the form $\ket{b} \propto \ket{\psi_0} - \ket{\psi_1}$. As proved in \cite{MeyerWong2015}, the energy gap at $\gamma_{c1}$ is $\Delta E = E_1 - E_0 = 4/M^{3/2}$, and so the system evolves from $\ket{s}$ to $\ket{b}$ in time $t_1 = \pi/\Delta E = \pi M^{3/2} / 4$.

In this first stage of the algorithm, we have moved probability from being uniformly distributed throughout the graph to the correct cluster (see Fig.~\ref{fig:simplex_weighted}). Now we want the probability to move within this cluster to the marked vertex. To do this, note from Fig.~\ref{fig:overlap_w1} that when $\gamma$ equals $\gamma_{c2} = 1/M = 0.001$, $\ket{b}$ is now half in the ground and third excited states, taking the form $\ket{b} \propto \ket{\psi_0} + \ket{\psi_3}$. At $\gamma_{c2}$, we also have the marked vertex $\ket{a} \propto \ket{\psi_0} - \ket{\psi_3}$ with an energy gap of $\Delta E = E_3 - E_0 = 2/\sqrt{M}$. So for the second stage of the algorithm, the system evolves from $\ket{b}$ to $\ket{a}$ in time $t_2 = \pi/\Delta E = \pi \sqrt{M} / 2$.

\begin{figure}
\begin{center}
	\includegraphics{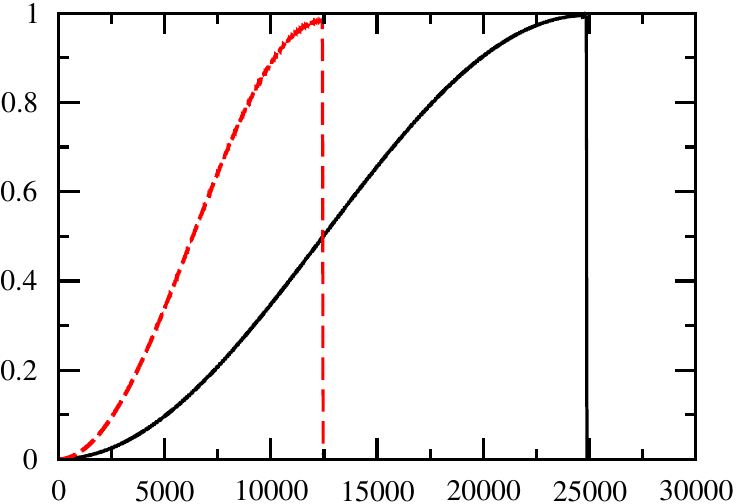}
	\caption{\label{fig:prob_time_b} Probability at $\ket{b}$ as a function of time for search on a simplex of complete graphs with $M = 1000$. The solid black curve is $w = 1$ (unweighted), and the dashed red curve is $w = 3$. Probability accumulates during the first stage of the algorithm and then it quickly leaves during the second stage (the sudden drop).}
\end{center}
\end{figure}

\begin{figure}
\begin{center}
	\includegraphics{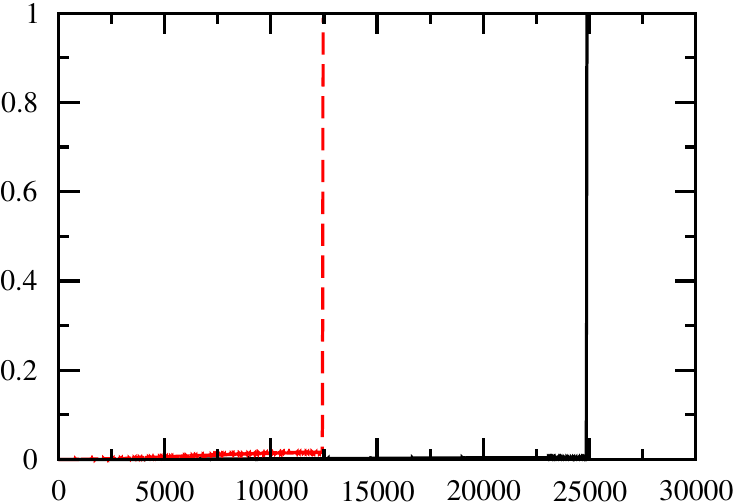}
	\caption{\label{fig:prob_time_a} Probability at $\ket{a}$ (\textit{i.e.}, the success probability) as a function of time for search on a simplex of complete graphs with $M = 1000$. The solid black curve is $w = 1$ (unweighted), and the dashed red curve is $w = 3$. During the second stage of the algorithm, the probability quickly accumulates (the sudden spike).}
\end{center}
\end{figure}

We can see each distinctive stage in Figs.~\ref{fig:prob_time_b} and \ref{fig:prob_time_a}, where we plot the probability at $\ket{b}$ and $\ket{a}$, respectively, as the system evolves. The unweighted ($w = 1$) case is the solid black curves. For the first stage of the algorithm, from $t = 0$ to $t_1 = \pi\,1000^{3/2} / 4 \approx 24836.471$, probability accumulates at $\ket{b}$. Then we switch to the second stage of the algorithm, which only lasts for time $t_2 = \pi \sqrt{1000} / 2 \approx 49.673$, where the probability quickly moves from $\ket{b}$ to $\ket{a}$, achieving the search.

The total runtime of the algorithm is the sum of the time for each stage: $t_1 + t_2 = \pi/\Delta E = \pi M^{3/2} / 4 + \pi \sqrt{M} / 2 = \Theta(M^{3/2}) = \Theta(N^{3/4})$. Clearly, the slow part of the algorithm is the first stage, where the probability is moving between clusters. The second stage is fast, where probability moves within the cluster containing the marked vertex. It is reasonable, then, that increasing the weights of the edges between clusters (the dotted edges in Fig.~\ref{fig:simplex_weighted}) can speed up this slow first stage. This is the idea of this paper.

Let us see how increasing the weights affects the algorithm. Fig.~\ref{fig:overlap_w3} shows the probability overlap plots with $w = 3$. From this, we see that the algorithm is still a two-stage evolution, except $\gamma_{c1}$ has shifted to the left. What is more striking, however, are the evolution plots in Figs.~\ref{fig:prob_time_b} and \ref{fig:prob_time_a}. Running each stage for the appropriate amount of time so that the probability moves from $\ket{s}$ to $\ket{b}$ to $\ket{a}$, we see that the runtime has been cut in half (actually, the first stage's runtime has been cut in half, and the second stage is the same).

So we indeed get a speedup by increasing the weights of the edges between the complete graph clusters. To prove this, and to show just how much of a speedup is possible, we use degenerate perturbation theory \cite{JMW2014}.


\section{Adjusted Degenerate Perturbation Theory}

\begin{figure}
\begin{center}
	\subfloat[]{
		\includegraphics{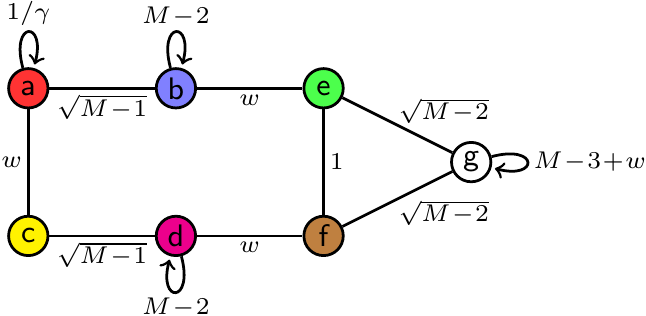}
		\label{fig:diagram_H} 
	}

	\subfloat[]{
		\includegraphics{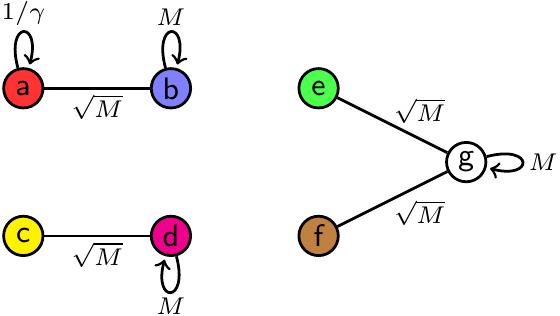}
		\label{fig:diagram_H_stage1} 
	}

	\subfloat[]{
		\includegraphics{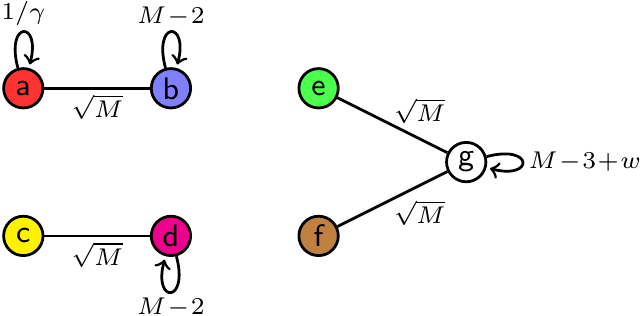}
		\label{fig:diagram_H_adjusted} 
	}

	\subfloat[]{
		\includegraphics{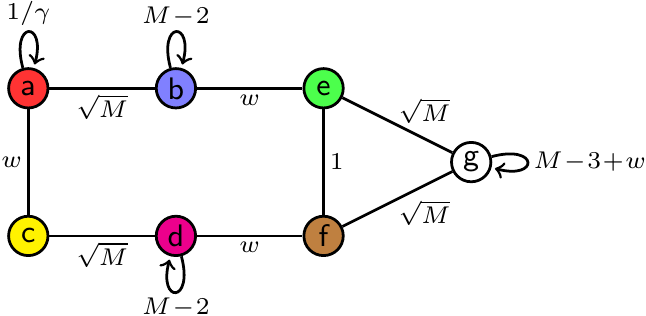}
		\label{fig:diagram_H_perturbed} 
	}

	\subfloat[]{
		\includegraphics{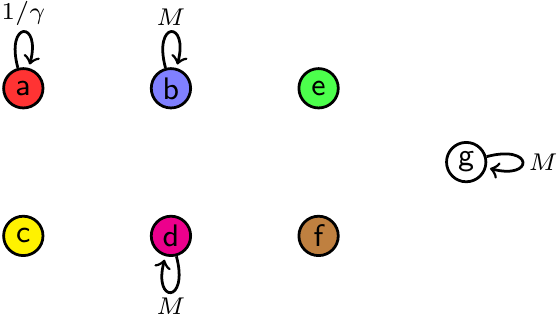}
		\label{fig:diagram_H_stage2} 
	}

	\caption{Apart from a factor of $-\gamma$, the (a) Hamiltonian for search on the weighted simplex of complete graphs, (b) typical first-stage leading-order terms, (c) adjusted first-stage leading-order terms, (d) first-stage leading-order terms plus a perturbation, and (e) second-stage leading-order terms. Note (b) is also the second-stage leading-order terms plus a perturbation.}
\end{center}
\end{figure}

To begin, we interpret the search Hamiltonian (in the 7-dimensional subspace) as an adjacency matrix, which we express diagrammatically \cite{Wong2014} in Fig.~\ref{fig:diagram_H}. Finding the eigenvalues and eigenvectors of this Hamiltonian is too complicated given all the connections between the basis states. To simplify it, we treat the Hamiltonian as a leading-order term $H^{(0)}$ plus a perturbation $H^{(1)}$. Assuming that $w$ scales less than $\sqrt{M}$ throughout this paper, we can drop it with all other terms that scale less than $\sqrt{M}$, yielding the leading-order Hamiltonian shown in Fig.~\ref{fig:diagram_H_stage1}. This is precisely the leading-order Hamiltonian for the unweighted ($w = 1$) graph, and its eigenvalues and eigenvectors are given in \cite{MeyerWong2015}. While we have made the leading-order Hamiltonian tractable, we have lost all dependence on the weight $w$.

We need to introduce some dependence on $w$ back, but not so much that the problem becomes intractable. To do this, consider the original simplex of complete graphs in Fig.~\ref{fig:simplex_weighted}. There are $M(M+1)/2$ edges with weight $w$. Some of them connect $a$ with $c$, some $b$ with $e$, some $d$ with $f$, and some $g$ with $g$. Just how many is shown in Table~\ref{table:weighted}. Thus the edges of weight $w$ are dominated by those connecting $g$ vertices with other $g$ vertices. So we add this back into the leading-order Hamiltonian.

\begin{table}
	\caption{\label{table:weighted}The number of edges of weight $w$ based on what kinds of vertices they connect.}
	\begin{ruledtabular}
	\begin{tabular}{cc}
		Connection & Number of Edges \\
		\colrule
		$a \sim c$ & $1$ \\
		$b \sim e$ & $M-1$ \\
		$d \sim f$ & $M-1$ \\
		$g \sim g$ & $\frac{(M-1)(M-2)}{2}$ \\
	\end{tabular}
	\end{ruledtabular}
\end{table}

For consistency, we should also do this for the edges with weight $1$. There is a total of $M(M+1)(M-1)/2$ such edges, and their connections are as shown in Table~\ref{table:unweighted}. These are dominated by $g \sim g$, followed by $b \sim b$, $d \sim d$, $e \sim g$, and $f \sim g$. Note that $e \sim g$ and $f \sim g$ are already included in Fig.~\ref{fig:diagram_H_stage1}, but we should add the other terms in.

\begin{table}
	\caption{\label{table:unweighted}The number of edges of weight $1$ based on what kinds of vertices they connect}
	\begin{ruledtabular}
	\begin{tabular}{cc}
		Connection & Number of Edges \\
		\colrule
		$a \sim b$ & $M-1$ \\
		$b \sim b$ & $\frac{(M-1)(M-2)}{2}$ \\
		$c \sim d$ & $M-1$ \\
		$d \sim d$ & $\frac{(M-1)(M-2)}{2}$ \\
		$e \sim f$ & $M-1$ \\
		$e \sim g$ & $(M-1)(M-2)$ \\
		$f \sim g$ & $(M-1)(M-2)$ \\
		$g \sim g$ & $\frac{(M-1)(M-2)(M-3)}{2}$ \\
	\end{tabular}
	\end{ruledtabular}
\end{table}

Including these additional contributions, we get the adjusted leading-order Hamiltonian in Fig.~\ref{fig:diagram_H_adjusted}. This now has dependence on $w$, but is still simple enough to solve analytically. In particular, there are two eigenstates of it that we are interested in:
\begin{gather*}
	u = \frac{1 + 2\gamma - M\gamma + \sqrt{R_u}}{2\sqrt{M}\gamma} \ket{a} + \ket{b} \\
	v = \frac{2 \sqrt{M}}{-3 + M + w + \sqrt{R_v}} \left( \ket{e} + \ket{f} \right) + \ket{g},
\end{gather*}
with respective eigenvalues
\begin{gather*}
	E_u = \frac{1-2 \gamma +M \gamma +\sqrt{R_u}}{2 \gamma} \\
	E_v = \frac{1}{2} \left(-3+M+w+\sqrt{R_v}\right),
\end{gather*}
where
\begin{gather*}
	R_u = 1 + 4\gamma - 2M\gamma + 4\gamma^2 + M^2\gamma^2 \\
	R_v = 9 + 2M + M^2 - 6w + 2Mw + w^2
\end{gather*}
are the radicands in the expressions. These two eigenstates are of interest because $\ket{u} \approx \ket{b}$ and $\ket{v} \approx \ket{g} \approx \ket{s}$ to leading order, and recall from Fig.~\ref{fig:overlap} that we want to choose $\gamma$ so that the ground and first excited states are each half $\ket{b}$ and half $\ket{s}$. If the eigenvalues $E_u$ and $E_v$ are nondegenerate, then even with the perturbation, $\ket{u}$ and $\ket{v}$ will remain approximate eigenstates. But when they are degenerate, then linear combinations of them
\[ \ket{\psi} = \alpha_u \ket{u} + \alpha_v \ket{v}, \]
will be eigenstates of the perturbed system, which is what we want. Setting their unperturbed eigenvalues $E_u$ and $E_v$ equal, solving for $\gamma$, and taking the leading-order term for large $M$, we get the critical $\gamma$ for the first stage of the algorithm:
\[ \gamma_{c1} = \left( 1 + \frac{1}{w} \right) \frac{1}{M}. \]
When $w = 1$, this yields the unweighted value of $2/M$, as expected from \cite{MeyerWong2015} and Fig.~\ref{fig:overlap_w1}. When $w = 3$, we get with $M = 1000$ that $\gamma_{c1} = (1 + 1/3)/1000 = 0.0013$, in agreement with Fig.~\ref{fig:overlap_w3}.

Now lets calculate the perturbed eigenstates. The perturbation $H^{(1)}$ restores terms of constant weight so that the we get Fig.~\ref{fig:diagram_H_perturbed}. This lifts the degeneracy so that, as described above, the eigenstates become $\alpha_u \ket{u} + \alpha_v \ket{v}$, and the coefficients can be found by solving
\[ \begin{pmatrix}
	H_{uu} & H_{uv} \\
	H_{vu} & H_{vv} \\
\end{pmatrix} \begin{pmatrix}
	\alpha_u \\
	\alpha_v \\
\end{pmatrix} = E \begin{pmatrix}
	\alpha_u \\
	\alpha_v \\
\end{pmatrix}, \]
where $H_{uv} = \bra{u} H^{(0)} + H^{(1)} \ket{v}$, etc. Calculating these matrix elements with $\gamma = \gamma_{c1}$, the terms $O(1/M^{3/2})$ are
\[ \begin{pmatrix}
	-\frac{1+w}{w} + \frac{1-w^2}{w} \frac{1}{M} & -\frac{1+w}{M^{3/2}} \\
	-\frac{1+w}{M^{3/2}} & -\frac{1+w}{w} + \frac{1-w^2}{w} \frac{1}{M} \\
\end{pmatrix} \begin{pmatrix}
	\alpha_u \\
	\alpha_v \\
\end{pmatrix} = E \begin{pmatrix}
	\alpha_u \\
	\alpha_v \\
\end{pmatrix}. \]
Solving this yields perturbed eigenstates
\[ \ket{\psi_{\pm}} = \frac{1}{\sqrt{2}} \begin{pmatrix} \pm 1 \\ 1 \end{pmatrix} = \frac{1}{\sqrt{2}} \left( \pm \ket{u} + \ket{v} \right) \approx \frac{1}{\sqrt{2}} \left( \pm \ket{b} + \ket{g} \right) \]
with corresponding eigenvalues
\[ E_\pm = -\frac{1+w}{w} + \frac{1-w^2}{wM} \mp \frac{1+w}{M^{3/2}}. \]
Thus to leading order, the system evolves from $\ket{s} \approx \ket{g}$ to $\ket{b}$ in time
\[ t_1 = \frac{\pi}{\Delta E} = \frac{\pi}{2(1+w)} M^{3/2}. \]
When $w = 1$, we get the unweighted graph's expected runtime of $\pi M^{3/2} / 4$ for the first stage of the algorithm. Clearly, as we increase $w$, the runtime is decreased. For example, when $w = 3$, the runtime is $\pi M^{3/2} / 8$, which is half the time of the unweighted graph; this agrees with Figs.~\ref{fig:prob_time_b} and \ref{fig:prob_time_a}, where the dashed red curve shows the probability at $\ket{b}$ and $\ket{a}$, respectively, as the system evolves with $w = 3$.

When $w$ scales larger than a constant, then the runtime scaling of $t_1$ is reduced. For example, if $w = \Theta(M^{1/4})$, then $t_1 = \Theta(M^{5/4}) = \Theta(N^{5/8})$. Recall we are assuming that $w$ scales less than $\sqrt{M}$ in order to employ our perturbative method, so $t_1$ can almost be lowered to $\Theta(M) = \Theta(\sqrt{N})$. As we will see later, there is another consideration to account for that yields this same limitation on $w$. But before focusing too much on $t_1$, we must determine how the weights affect the second stage of the algorithm.


\section{Stage 2}

Let us now analyze how the weights affect the second stage of the algorithm, where probability moves from $\ket{b}$ to $\ket{a}$. From Fig.~\ref{fig:overlap}, the critical $\gamma$ for the second stage of the algorithm is unchanged from its unweighted value of $\gamma_{c2} = 1/M$. To prove this, we again use degenerate perturbation theory. This time, there is no adjustment needed as was in the first stage to reintroduce dependence on $w$, so the results from \cite{MeyerWong2015} and \cite{Wong2014} carry over directly, and which we summarize here.

For the leading-order Hamiltonina $H^{(0)}$, we drop the terms that scale less than $M$, yielding Fig.~\ref{fig:diagram_H_stage2}. From this, we see that $\ket{a}$ and $\ket{b}$ are eigenstates of $H^{(0)}$ with respective eigenvalues $-1$ and $-\gamma M$. If these eigenvalues are nondegenerate, then even with the perturbation, $\ket{b}$ will be an approximate eigenstate of the Hamiltonian, so the system will not evolve from $\ket{b}$ apart from a global, unobservable phase. To make the system evolve, we equate the eigenvalues to create a degeneracy, yielding the critical $\gamma$ for the second stage of the algorithm:
\[ \gamma_{c2} = \frac{1}{M}. \]
Then the perturbation $H^{(1)}$, which restores edges of weight $\Theta(\sqrt{M})$ so that we have Fig.~\ref{fig:diagram_H_stage1}, cause the eigenstates to become
\[ \alpha_a \ket{a} + \alpha_b \ket{b}, \]
where the coefficients can be found by solving
\[ \begin{pmatrix}
	H_{aa} & H_{ab} \\
	H_{ba} & H_{bb} \\
\end{pmatrix} \begin{pmatrix}
	\alpha_a \\
	\alpha_b \\
\end{pmatrix} = E \begin{pmatrix}
	\alpha_a \\
	\alpha_b \\
\end{pmatrix}, \]
where $H_{ab} = \bra{a} H^{(0)} + H^{(1)} \ket{b}$, etc. With $\gamma = \gamma_{c2}$, this is
\[ \begin{pmatrix}
	-1 & \frac{-1}{\sqrt{M}} \\
	\frac{-1}{\sqrt{M}} & -1 \\
\end{pmatrix} \begin{pmatrix}
	\alpha_a \\
	\alpha_b \\
\end{pmatrix} = E \begin{pmatrix}
	\alpha_a \\
	\alpha_b \\
\end{pmatrix}. \]
Solving this yields perturbed eigenstates and eigenvalues
\[ \frac{1}{\sqrt{2}} \left( \pm \ket{a} + \ket{b} \right), \quad -1 \mp \frac{1}{\sqrt{M}}. \]
Thus during the second stage of the algorithm, the system evolves from $\ket{b}$ to $\ket{a}$ in time
\[ t_2 = \frac{\pi}{\Delta E} = \frac{\pi \sqrt{M}}{2}. \]
So adding weights to the graph does not change the runtime of the second stage of the algorithm. Thus we can speed up the first stage of the algorithm by making the graph weighted without harming the second stage of the algorithm, and the total runtime is
\[ t_1 + t_2 = \frac{\pi M^{3/2}}{2(1+w)} + \frac{\pi \sqrt{M}}{2}. \]
This agrees with Figs.~\ref{fig:prob_time_b} and \ref{fig:prob_time_a}, where for $w = 3$ and $M = 1000$, the system evolves for time $t_1 = \pi 1000^{3/2} / 2(1+3) \approx 12418.235$, and then for an additional time of $t_2 = \pi \sqrt{1000}/2 \approx 49.673$.

Thus to reduce the overall runtime, we want to increase $w$ to be as large as possible. Recall that these results assume that $w$ scales less than $\sqrt{M}$, so we can almost decrease the overall runtime to $\Theta(M) = \Theta(\sqrt{N})$. But for this to be possible, there is one more matter to consider regarding the critical $\gamma$'s.


\section{Convergence of Critical Gammas}

Recall the critical $\gamma$'s for the first and second stages of the algorithm:
\[ \gamma_{c1} = \left( 1 + \frac{1}{w} \right) \frac{1}{M}, \quad \gamma_{c2} = \frac{1}{M}. \]
Then as $w$ increases, $\gamma_{c1}$ converges to $\gamma_{c2}$. This can also be seen in Fig.~\ref{fig:overlap}; as $w$ increases, the overlap crossing at $\gamma_{c1}$ moves to the left until it collides with the stationary crossing at $\gamma_{c2}$. This causes the distinct evolution of the two stages to meld together in a way that is unpredictable by our current method of degenerate perturbation theory.

So how big can $w$ be such that the two-stage algorithm holds? This equates to finding the ``width'' around each critical $\gamma$ in which the algorithm evolves correctly. For example, for the complete graph of $N$ vertices, an explicit calculation in \cite{Wong2015c} shows that when $\gamma$ is within $O(1/N^{3/2})$ of its critical value of $1/N$, then the system searches in Grover's $\Theta(\sqrt{N})$ time. When $\gamma$ is outside of this region, the system asymptotically starts in an eigenstate and fails to evolve apart from a global, unobservable phase. So there is a region around the critical $\gamma$ in which the algorithm is relevant, and outside of which it is not. For our weighted graph, we want the $\gamma_c$'s to be outside of each other's region of influence.

To find the relevant region around each $\gamma_c$, an explicit calculation as in \cite{Wong2015c} is prohibitive. Instead, we introduce a new approach to finding such precision bounds on how close $\gamma$ must be to its critical value using degenerate perturbation theory. To demonstrate it, let us start with the second stage, which is easier and will yield the relevant region. Recall from the last section that $\ket{a}$ and $\ket{b}$ are eigenstates of the leading-order Hamiltonian in Fig.~\ref{fig:diagram_H_stage2} with respective eigenvalues $-1$ and $-\gamma M$. Then if $\gamma$ is $\epsilon$ away from its critical value, \textit{i.e.}, $\gamma = \gamma_{c2} + \epsilon = 1/M + \epsilon$, then $\ket{b}$'s unperturbed eigenvalue becomes $-1 - \epsilon M$. If $\epsilon$ is small so that $\ket{a}$ and $\ket{b}$ are near-degenerate, then with the perturbation, the eigenstates become linear combinations $\alpha_a \ket{a} + \alpha_b \ket{b}$, and we solve an eigenvalue problem for the coefficients. The eigenvalue problem contains a term $H_{bb} = \langle b | H^{(0)} + H^{(1)} | b \rangle$, which contains a term scaling as $\epsilon M$ that is the leading-order term in $\epsilon$. For the system to evolve with the correct scaling, $\epsilon M$ must scale no greater than the energy gap $\Theta(1/\sqrt{M})$. Thus $\epsilon = O(1/M^{3/2})$. This gives the precision with which $\gamma$ must equal its second critical value.

Since the first stage's energy gap $\Theta(w/M^{3/2})$ is smaller than the second stage's $\Theta(1/\sqrt{M})$, the first stage has a smaller range of $\gamma$'s that will work for it. This can also be seen in Fig.~\ref{fig:overlap_w3}, where $\gamma_{c1} = 0.0013$ has a thinner crossing than $\gamma_{c2} = 0.001$'s crossing. Thus when the $\gamma_c$'s ``collide,'' it is when they are within $O(1/M^{3/2})$ of each other, having entered the second stage's region of influence.

Back to our original question of how big $w$ can be before the first stage ``collides'' with the second stage. We want
\[ \gamma_{c1} = \frac{1}{M} + \frac{1}{wM} = \gamma_{c2} + \frac{1}{wM} \]
to stay outside of the range of the second stage. Taking $1/wM$ to be $\epsilon$ in the previous calculation, we want it to scale bigger than $1/M^{3/2}$, which yields
\[ w = o(\sqrt{M}). \]
That is, the weight must scale less than $\sqrt{M}$ for the two stages of the algorithm to be distinct. Since this is the assumption throughout our paper for the perturbative calculations to work, our algorithm works in all values of $w$ considered in this work, and it allows the runtime to be reduced to nearly $\Theta(\sqrt{N})$.


\section{Energy and Connectivity}

We end with some remarks about the energy usage of the algorithm, as well as the connectivity of the weighted graph. First regarding energy, the operator norm of the Hamiltonian gives some sense of the energy being used. By making the graph weighted, we have only changed $\gamma$ and the adjacency matrix while leaving the oracle term $\ketbra{a}{a}$ alone. The operator norm of the adjacency matrix is $M + w - 1 = \Theta(M)$, and both the critical $\gamma$'s scale as $\Theta(1/M)$. Thus the operator norm of $\gamma A$ is $\Theta(1)$, and so to leading order, the search algorithm on the weighted graph does not use more energy than on the unweighted graph.

Now for connectivity. Since the (unweighted) simplex of complete graphs was first introduced for quantum computing as a graph with high connectivity, but on which search is slow \cite{MeyerWong2015}, we comment on the connectivity of our weighted version. The vertex connectivity is unchanged, since $M$ vertices must be removed to disconnect the graph. Thus it does not capture the weights and is a poor indicator of connectivity for weighted graphs. The edge connectivity is also unsatisfactory. While $M$ edges must be removed, should the weights of the edges be accounted for? For example, a cluster can be disconnected from the rest of the graph by removing the $M$ edges of weight $w$ connecting to the cluster. Or a single vertex can be disconnected by removing its $M$ edges, of which $M-1$ have weight $1$ and one has weight $w$. As such, it is better to consider algebraic connectivity \cite{Fiedler1973} for weighted graphs, which is the second smallest eigenvalue of the graph Laplacian $L = D - A$, where $D$ is a diagonal matrix with the degree of each vertex. For the weighted simplex of complete graphs, it is
\[ \lambda_1 = \frac{1}{2} \left( M + 2w - \sqrt{M^2 - 4w + 4w^2} \right) \approx w. \]
Clearly, making the weight $w$ larger increases the algebraic connectivity, as expected. We can further improve this with a normalized algebraic connectivity \cite{Chung1997}, which for a regular graph is simply its algebraic connectivity divided by its degree, in this case yielding roughly $w / M$. As expected, this connectivity is higher when the weight $w$ increases. When $w$ almost scales as $\sqrt{M}$ so that the runtime is nearly reduced to $\Theta(\sqrt{N})$, the algebraic connectivity is almost that of an eight-dimensional cubic lattice, on which search is fast \cite{CG2004}.

This does not contradict the conclusion of \cite{MeyerWong2015}, of course. That the weighted simplex of complete graphs has high connectivity and yields fast search does not take away from the counterexample of the unweighted graph searching slower than its connectivity would suggest. Furthermore, by changing the weights, we have changed the graph.


\section{Conclusion}

We have modified the simplex of complete graphs to have different weights on the edges between clusters from the edges within clusters. This defines a reasonable search structure that is vertex-transitive, and whose equal superposition over the vertices is an eigenstate of its adjacency matrix. By adjusting degenerate perturbation theory, we proved that the runtime on this weighted graph can be reduced from the unweighted $\Theta(N^{3/4})$ to nearly $\Theta(\sqrt{N})$. Thus we have novelly demonstrated that faster quantum search is possible on a weighted graph.

It is very possible that our work can be extended to fully reduce the runtime to Grover's $\Theta(\sqrt{N})$. For example, we have left open the case when $w$ scales greater than or equal to $\sqrt{M}$. One could also consider changing the weights differently, moving away from our model where edges between clusters have one weight while edges within clusters have another. Finally, different graphs could be considered altogether.


\begin{acknowledgments}
	Thanks to a referee of \cite{Wong2015a} for suggesting search on the weighted simplex of complete graphs. This work was supported by the European Union Seventh Framework Programme (FP7/2007-2013) under the QALGO (Grant Agreement No.~600700) project, and the ERC Advanced Grant MQC.
\end{acknowledgments}


\bibliography{refs}

\end{document}